# Theoretical and Experimental Studies of Schottky Diodes That Use Aligned Arrays of Single Walled Carbon Nanotubes

Xinning Ho[1+], Lina Ye[1,2+], Slava V. Rotkin[3+*], Xu Xie[1], Frank Du[1], Simon Dunham[1], Jana Zaumseil[4] and John A. Rogers[1,5,6*]

*[1]Departments of Materials Science and Engineering, Beckman Institute, and Frederick Seitz Materials Research Laboratory, University of Illinois at Urbana-Champaign, 1304 West Green Street, Urbana, Illinois 61801*

*[2]Department of Chemistry, University of Science and Technology of China, Hefei 230026, P.R.China*

*[3]Department of Physics, Lehigh University, Bethlehem, Pennsylvania 18015 and Centre for Advanced Materials and Nanotechnology, Lehigh University, Bethlehem, Pennsylvania 18015*

*[4]Center for Nanoscale Materials, Argonne National Laboratory, Argonne, Illinois 60439*

*[5]Department of Chemistry, University of Illinois at Urbana-Champaign, Urbana Illinois 61801*

*[6]Department of Electrical and Computer Engineering, Mechanical Science and Engineering, University of Illinois at Urbana-Champaign, Urbana, Illinois 61801*

*+ These authors contributed equally to this work.*

***To whom correspondence should be addressed, jrogers@uiuc.edu ; rotkin@lehigh.edu***

**Abstract.** We present theoretical and experimental studies of Schottky diodes that use aligned arrays of single walled carbon nanotubes. A simple physical model, taking into account the basic physics of current rectification, can adequately describe the single-tube and array devices. We show that for as grown array diodes, the rectification ratio, defined by the maximum-to-minimum-current-ratio, is low due to the presence of m-SWNT shunts. These tubes can be eliminated in a single voltage sweep resulting in a high rectification array device. Further analysis also shows that the channel resistance, and not the intrinsic nanotube diode properties, limits the rectification in devices with channel length up to ten micrometer.

Since the earliest days of work on semiconductor devices, diodes have played critically important roles. Theories of p-n Schottky diodes [1-2] laid the foundations for understanding bi-polar transistor operation and contact phenomena at the metal/semiconductor interface. Even though the diode itself is not a main element of modern digital electronics, the physics of the diode structure is essential for many applications, including in optoelectronics [3]. Nanoscale diodes have been already demonstrated with carbon-based nano-materials, such as graphene and individual nanotubes [4-14]. The work presented here focuses on diode structures made of parallel nanotube arrays, their rectification properties and the physics of their electronic transport. The array format is advantageous because they deliver much larger currents than a single-tube device and have less noise, enabling them to operate at high-frequency as we have demonstrated extensively in transistors, amplifiers and even fully integrated transistor radios [15-18]. In addition, arrays provide a natural path toward large scale integration, in which the spatial position and electronic properties of any given tube in the array are not critically important; the large numbers of tubes that contribute to operation of a given device yield statistical averaging effects that can provide good device-to-device uniformity in properties. In currently achievable, as-grown arrays, both semiconducting and metallic species are present, thus making the physics of diode operation more complex than that given by the textbook equations [19]. In this Letter, we combine experiments with a compact model for device behavior to reveal key aspects. Below we present a clear physical interpretation for the transport experiments with nanotube array diodes as well as for similarly fabricated individual tube diodes. This outcome allows us to extract the average device parameters and correlate them to the physical properties of individual tubes.

We start by considering Schottky diodes based on single semiconducting-SWNTs (s-SWNTs) and resistors based on single metallic-SWNTs (m-SWNTs). Figure 1a shows a schematic illustration of such devices. For these cases, as well as the array devices described next, we grew perfectly aligned parallel arrays of SWNTs via chemical vapor deposition (CVD) on a ST (stable temperature) cut quartz, using procedures described elsewhere [20,21]. Electrodes defined by photolithography and liftoff were deposited directly on the arrays. One electrode was Pd(30nm)/Ti(1nm), providing the Ohmic contact [22-24] and the other electrode was Al(30nm)/Ca(3nm), providing the Schottky contact [9, 25] to the s-SWNTs. Since the thickness of deposited Ti layer was smaller than a typical SWNT diameter, the SWNTs had actual contacts to the Pd-bulk with the work function ~5.1eV [22-24], making p-contacts. On the other hand, the Ca layer, with the work function ~2.9eV [9,25], was sufficiently thick to coat completely the SWNTs and to make n-type contact. The channel lengths and channel widths were ~10μm and ~15μm, respectively. Narrow stripes (~1μm wide) of photoresist (AZ5214) defined via photolithography protected selected areas of SWNTs before we placed the substrate into a Reactive Ion Etcher (RIE) to oxygen etch unwanted SWNTs. This process increases the chance of obtaining a single SWNT per device by decreasing the number of SWNTs that bridge the metal electrodes and it also electrically isolates neighboring devices. A Scanning Electron Microscope (SEM) was used to determine the number of tubes in each device, after electrical characterization using a Parameter Analyzer (Agilent 4156A).

Parts b and c of figure 1 show typical current-voltage (I-V) curves of single m-SWNT resistors and single s-SWNT diodes, respectively, measured with the Al/Ca electrode grounded and the Pd electrode swept between ±2V. We note that though performance generally degrades over time for open diodes in air, when encapsulated with a layer of PMMA or stored in a glovebox the devices have long lifetimes. The I-V curves of the single m-SWNT resistors are linear while those of the single s-SWNT diodes display rectification at reverse bias, as might be expected. In an ideal diode structure, thermionic current, $I_d$, follows a simple exponential dependence on the drive voltage, $V_d$:

$I_d = I_o(\exp[eV_d / nk_BT] - 1)$ where $I_o$ and $n$ are the reverse saturation current and non-ideality factor respectively, $e$ and $k_B$ are electron charge and Boltzmann constant respectively and $k_BT/e$ is the temperature in Volts. None of our devices shows this simple I-V curve, which is consistent with other studies [5-6, 10-13, 26]. However, we propose a physical model that can adequately describe the behavior. Our devices have channel lengths that exceed the length of the diode junction itself. Thus, the total voltage drop across the whole device is distributed between the junction and the rest of the channel. In addition, we find that a non-negligible current can flow at reverse bias. In multiple tube diodes this current is due to the shunt represented by the m-SWNTs. Since all tubes contribute to the total current in parallel, we derive the analytical expression for the I-V curve as:

$$I_d = \frac{V_d}{R_m} + I_o\left(\exp\left[\frac{e(V_d - I_dR_c)}{nk_BT}\right] - 1\right) = \frac{V_d}{R_m} + \frac{nk_BT}{eR_c}\varpi\left(\frac{eI_oR_c}{nk_BT}\exp\left[\frac{e(V_d + I_oR_c)}{nk_BT}\right]\right) - I_o \quad (1)$$

Here the expression in the middle is still an implicit function of $I_d$ and must be further solved for $V_d$. Because the solution is not available in elementary functions, we apply the product-log function to the expression on the right hand side [27, 28]: $y = \varpi(x)$ defined such that $x = y\log(y)$. The first term $V_d/R_m$ is due to the metallic shunt, with $R_m$ corresponding to a characteristic leakage resistance. $R_c$ represents a characteristic resistance of the physical diode. $I_dR_c$ is therefore the voltage drop at the SWNT channel and electrodes, except for that of the junction itself. The junction is characterized by the single parameter $I_o$. Thus the model has three fitting parameters, besides the ratio $(n\ k_BT/e)$ which we assume is fixed at a given temperature $T=300K$ and $n=1$ (thus we neglect the trap recombination below). Eq.(1) corresponds to the equivalent circuit shown in figure 1d.

The slope at reverse bias is determined by the $R_m$ term. $I_o$ can be dropped here because of its negligible numerical value. This result implies that in our devices the leakage is not due to the thermionic current through the diode junction itself (i.e., the upper path in the equivalent circuit of figure 1d is shut-off at $V_d << 0$. We provide full proof next.) For array devices, the slope of the reverse bias wing is very close to linear and unambiguously yields $R_m$ for m-SWNTs. This resistance is an order of magnitude higher than that for Pd-Pd contacted field effect transistors (FETs) [29] as obtained from known contact resistance and channel resistivity of similarly grown m-SWNTs, the measured channel length and width and estimated number of tubes in the array diode. We attribute this difference to the lower quality of the Ca contact, due at least partly to its poor wetting properties on SWNTs [9, 25] and to the higher drain bias applied in the diodes here (up to -2V) compared to that (-0.01 V) used in previous work on related transistor devices. [29]

At high positive bias, the diode structure is fully open and the physics is also simple: the equivalent circuit contains only $R_m$ and $R_c$. Theoretically, the product-log function saturates at large arguments: $\varpi(x) \to \log(x) + \ldots$. As a result, Eq.(1) reduces to: $I_d \approx \frac{V_d}{R_m} + \frac{V_d}{R_c} - \frac{nk_BT}{eR_c}\log\left(\frac{V_d + I_oR_c}{I_oR_c}\right)$. Thus, the linear slope of the high-bias I-V curve gives us the total device conductance: $\frac{1}{R_{tot}} = \frac{1}{R_m} + \frac{1}{R_c}$. The current cut-off, the point where the linear part of the I-V curve crosses the ordinate axis, with the logarithmic accuracy, is given by: $\frac{nk_BT}{eR_c}\log\left(\frac{nk_BT}{eI_oR_c}\right)$. In this manner, $I_o$ and $R_c$ can be extracted.

For example, figure 1e shows the current for single s-SWNT diodes, plotted on a log-scale. The measured leakage current at reverse bias allows us to determine $R_m \sim 1\text{-}40$ GΩ; the linear currents at high forward bias yield $R_c \sim 1\text{-}50$ MΩ, with $I_o \sim$fA or smaller. This observation proves post factum that the leakage current is not due to $I_o$. We note that at large $V_d$ the diode junction is open and has almost zero resistance. Thus, very little voltage drops at the junction $V_c \sim (n\, k_B T/e) \log(n\, k_B T/eI_oR_c) \sim 0.1\text{-}0.3$ V. The rest of the drop is due to Ohmic losses at the s-SWNT channel and electrodes, except for that of the junction itself. The extracted model parameter $R_c$ is about an order of magnitude higher than similarly fabricated FET devices [29] which we speculate is due to the lower quality of the Ca contact and the higher drain bias applied in the diodes here compared to that applied in previously studied transistors [29], similar to the case of m-SWNT devices. At small bias $V \ll V_c$, the junction resistance becomes very high, on the order of $nk_BT/eI_o \sim$TΩ. In this regime, almost all voltage drops at the junction, and not in the channel or electrodes. The resistance increases further at reverse bias, which must essentially shut off thermionic conduction through the ideal diode. In a real system we always observe leakage. The origin of the leakage for SWNT devices is unknown. Zener tunneling through the Schottky contact and thermal generation in the field region of the small bandgap SWNTs (<1eV) could explain the leakage current [6, 8]. Recently Schottky barrier tunneling currents were found to explain the reverse bias transport in ZnO nanowires [30]. Details of the modeling of the single s-SWNT diodes and m-SWNT resistors can be found in the supplementary information.

Figure 1f shows the rectification (i.e. ON-current at maximum positive bias of 2V divided by the absolute value of the current at minimum bias of -2V) as a function of the current at 2V for the various single m-SWNT resistors and single s-SWNT diodes. The rectification of the single m-SWNT resistors is about 1 while that of the single s-SWNT diodes could be as large as $10^4$. We emphasize that according to our analysis, the rectification is not limited by the internal physical properties of the material. Instead, the ON-current and thus the rectification are substantially limited by the channel resistance, being proportional to the channel length, similar to the ON-current in SWNT array and single-tube FET devices. We propose that short channel devices should achieve much better rectification due to lower resistance.

Having established physical parameters of single SWNT devices, we proceed to analysis of the arrays. Figure 2a,b show a schematic illustration of a Schottky diode based on perfectly aligned arrays of SWNTs and an SEM image of an array representative of the type used here. These array diodes were fabricated in a manner similar to the single SWNT devices mentioned earlier but the patterned photoresist covered the entire diodes to protect all SWNTs within the diodes during the etching process. The channel lengths and channel widths of these devices were ~10μm and ~250μm, respectively. The densities of the arrays (measured in SWNTs per micrometer of lateral distance across the channel) were 1 ± 0.5 SWNTs/μm, as determined by the average of SEM measurements at various spots across the surface of the quartz substrate. As a result, if we assume that the ratio of m-SWNTs to s-SWNTs is 1:2, then there are approximately ~83 m-SWNTs and ~167 s-SWNTs in each array SWNT diode. Figure 2c shows the diameter distribution of the SWNTs in the arrays, as determined by AFM measurements. The diameter of the SWNTs ranges from ~0.5nm to ~1.7 nm with single counts for tubes with diameters up to 4.8nm which we assume are small bundles of SWNTs. The majority of the SWNTs have diameters between 1.0 and 1.2nm.

Array SWNT diodes were measured with the Al/Ca electrode grounded and the Pd electrode swept between ±2V. A small rectification (~1.5) is observed in these array SWNT diodes. This result is consistent with the significant population of SWNTs that are m-SWNT and act as shunt channels.

Assuming that most of the leakage in reverse bias is due to these m-SWNT shunts, and then extracting the s-SWNT channel resistance at large forward bias as explained before, we can fit the array data. Next we compare the currents flowing through the m- and s-SWNTs throughout the range of biases. At reverse bias, m-SWNTs always dominate while at forward bias, we observe two cases (as shown in figure 2d,e). In figure 2d, the s-SWNTs resistance is approximately half of the m-SWNTs, which is reasonable assuming the ratio of s-SWNTs to m-SWNTs to be about 2:1. On the other hand, in figure 2e, these resistances are about the same, which may be due to stronger scattering at the contacts and/or in the channels of s-SWNTs. [31-33] Assuming that our single-SWNT device measurements have sampled sufficiently the random distribution of the SWNT channels in the array devices, we compare in figure 2f the experimental array IV curve and the one composed from an average m-SWNT and average s-SWNT IV curves weighted with their abundances in the arrays. In the Supplementary Materials we provide statistical analysis of the single-SWNT device data.

We can increase the rectification in the array devices by electrically breaking down most of the m-SWNTs. After applying $V_d$~30V, the I-V curve of the device shown in figure 3a demonstrates irreversible changes (compare to figure 3b before breakdown). Analysis of the device before and after the high-bias sweep indicates that current contributed by both m-SWNTs and s-SWNTs have decreased. However, the metallic-shunt resistance has increased much more significantly as shown in figure 3c. Thus, we speculate that we were able to burn preferentially m-SWNTs, to yield an array diode with good rectification. $\frac{R_m}{R_c}+1$, which also corresponds to the rectification ratio, increased from 1.6 to 29.3 after breakdown. After we applied even more significant bias sweep up to 50V, the s-SWNT channels also break down, and both resistances further increased to yield low ratios (figure 3c).

In conclusion, we present theoretical and experimental studies of diodes based on parallel arrays of SWNTs. A simple physical model takes into account basic physics of current rectification and explains the data. Our analysis is equally applicable to single-tube and array devices though we stress that some aspects of the charge carrier transport need further study, for example, the origin of the SWNT leakage current requires special attention. We show that for as grown array diodes, the rectification ratio, that is the maximum-to-minimum-current-ratio, is low due to the presence of m-SWNT shunts. These tubes can be eliminated in a single voltage sweep resulting in a high rectification array device. Further analysis shows that the channel resistance, and not the intrinsic nanotube diode properties, ultimately limits the rectification. Shorter devices may demonstrate even better performance, with some potential to serve in ultra-miniaturized circuits.

**Acknowledgment**. We thank T. Banks and B. Sankaran for help with processing. This work was carried out in part in the Frederick Seitz Materials Research Laboratory Central Facilities, University of Illinois, which are partially supported by the U.S. Department of Energy under Grants DE-FG02-07ER46453 and DE-FG02-07ER46471. X.H. acknowledges fellowship support from ASTAR (Singapore).

# Electronic Supplementary Material

## Theoretical and Experimental Studies of Schottky Diodes That Use Aligned Arrays of Single Walled Carbon Nanotubes

Xinning Ho, Lina Ye, Slava V. Rotkin, Xu Xie, Frank Du,
Simon Dunham, Jana Zaumseil and John A. Rogers

### Additional details on rectification ratio

Details of the magnitude of the current at $V_d = \pm 2V$ of all m-SWNTs and s-SWNTs can be found in supplementary figure S1.

### Additional details on fitting single-tube diodes

Using the modeling fit described in the main text (the equivalent circuit shown in figure 1(d) of the main text) we modeled single s-SWNT diodes as shown in supplementary figure S2. Top row shows two representative devices with a good fit to the model, described in the text. Some of fabricated devices showed less ideal fit to the model (bottom row of Fig. S2). Non-idealities, due possibly to the contacts, the presence of defects in some s-SWNTs and series resistance that could be drain bias dependent, might explain the discrepancies with the theoretical fits.

### Additional details on averaging procedure for array diodes

Figure 2(f) of main text shows the current contributed by s-SWNTs and m-SWNTs in an array SWNT diode, simulated from the current of single SWNT devices. Based on SWNT densities measurements (~ 1 ± 0.5 SWNTs/μm) from the SEM and the channel width (~ 250μm) of the array SWNT diodes, we estimated there to be ~250 SWNTs in each array SWNT diode. Assuming that the ratio of m-SWNTs to s-SWNTs is 1:2 in the array SWNT diode, there are approximately ~83 m-SWNTs and ~167 s-SWNTs in the array SWNT diode. Assuming the sampling was sufficient in our single-tube device measurements, we can interpolate the array data using the data from the single m-SWNT resistors and single s-SWNT diodes. Multiplying the averaged current of the single m-SWNT resistors by a factor of 83, we obtain the simulated current contributed by m-SWNTs in the array SWNT diode. Similarly, we multiply the averaged current of the single s-SWNT diodes by a factor of 167 to obtain the simulated current contributed by s-SWNTs in the array SWNT diode. Summing up the currents of the m-SWNTs and s-SWNTs, we obtain the total current of the array SWNT diode simulated from measurements based on single SWNT devices. This is found to be very similar to the measured I-V curve of an array SWNT diode in figure 2(f).

### Additional details on statistics of fit parameters for single-tube diodes

We analyzed the statistics of the measured single-tube devices. Supplementary figure S3(a) shows the probability distribution function (PDF) for the resistance of the m-SWNTs. The PDF has been determined from the data by computing the intervals between p-quantiles of the random variable (being the resistance of m-SWNTs in this case). To calculate the p-quantiles we use the mode-based estimate algorithm in Wolfram©Mathematica 7.0. The quantiles are computed subsequently at probabilities $0 \leq p \leq 1$ with sufficiently small increment of the probability, $\delta p << p$. We verify that the result is independent of the choice of $\delta p$.

Since the IV curves for m-SWNTs are slightly asymmetric, we measure the linear slope at both the forward (black curve) and reverse bias (red curve). Two peaks in the PDF may correspond to truly metallic (armchair) SWNTs and narrow band pseudo-metallic (chiral) SWNTs.

Similar PDFs have been obtained for s-SWNT diodes using the fit Eq.(1) of the main text (Fig. S3(b) and (c) and (d)). The PDFs for all three parameters $R_m$, $R_c$ and $I_o$, show sharp decrease with increasing values of the parameters. On the figures we plot PDFs in the log scale which indicates possible exponential decrease of the probability density. More statistical data is needed though to make more quantitative conclusions.

Supplementary figure S1

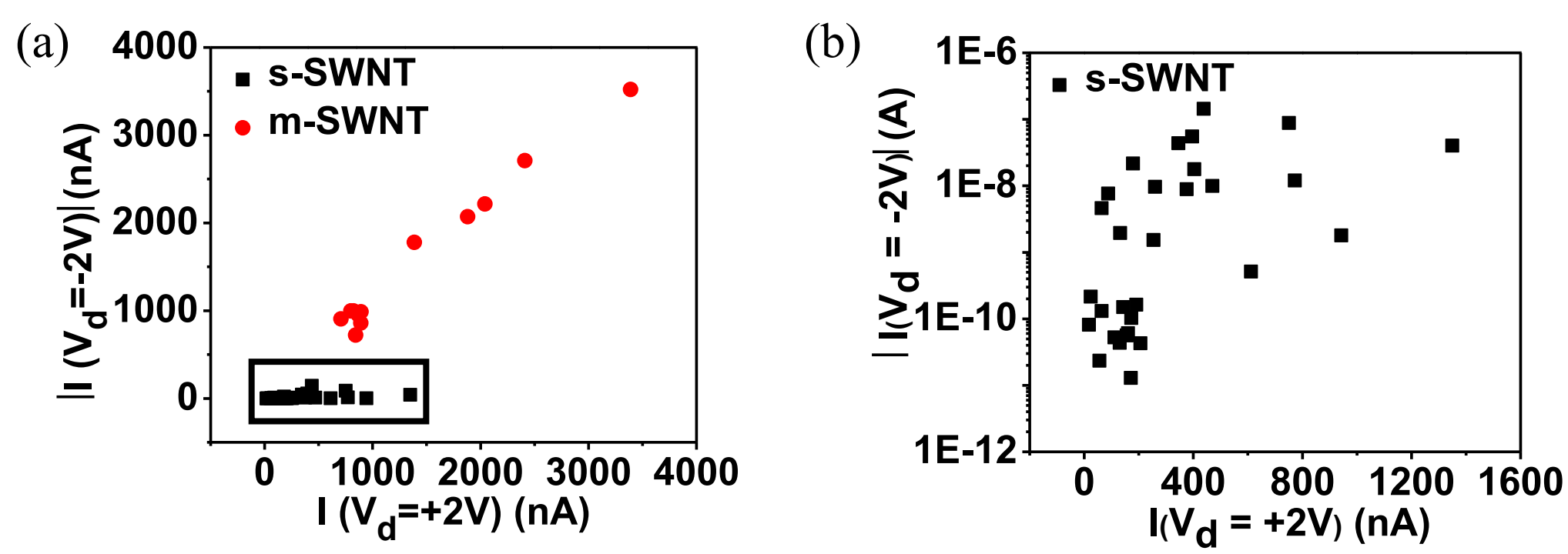

(a) Absolute value of current at $V_d$=-2V versus current at $V_d$=+2V graphs for single s-SWNT diodes (black squares) and single m-SWNT resistors (red circles).
(b) Same plot as in the boxed up region in part (a) (i.e. the s-SWNT diodes) shown in log 10 scale.

Supplementary figure S2

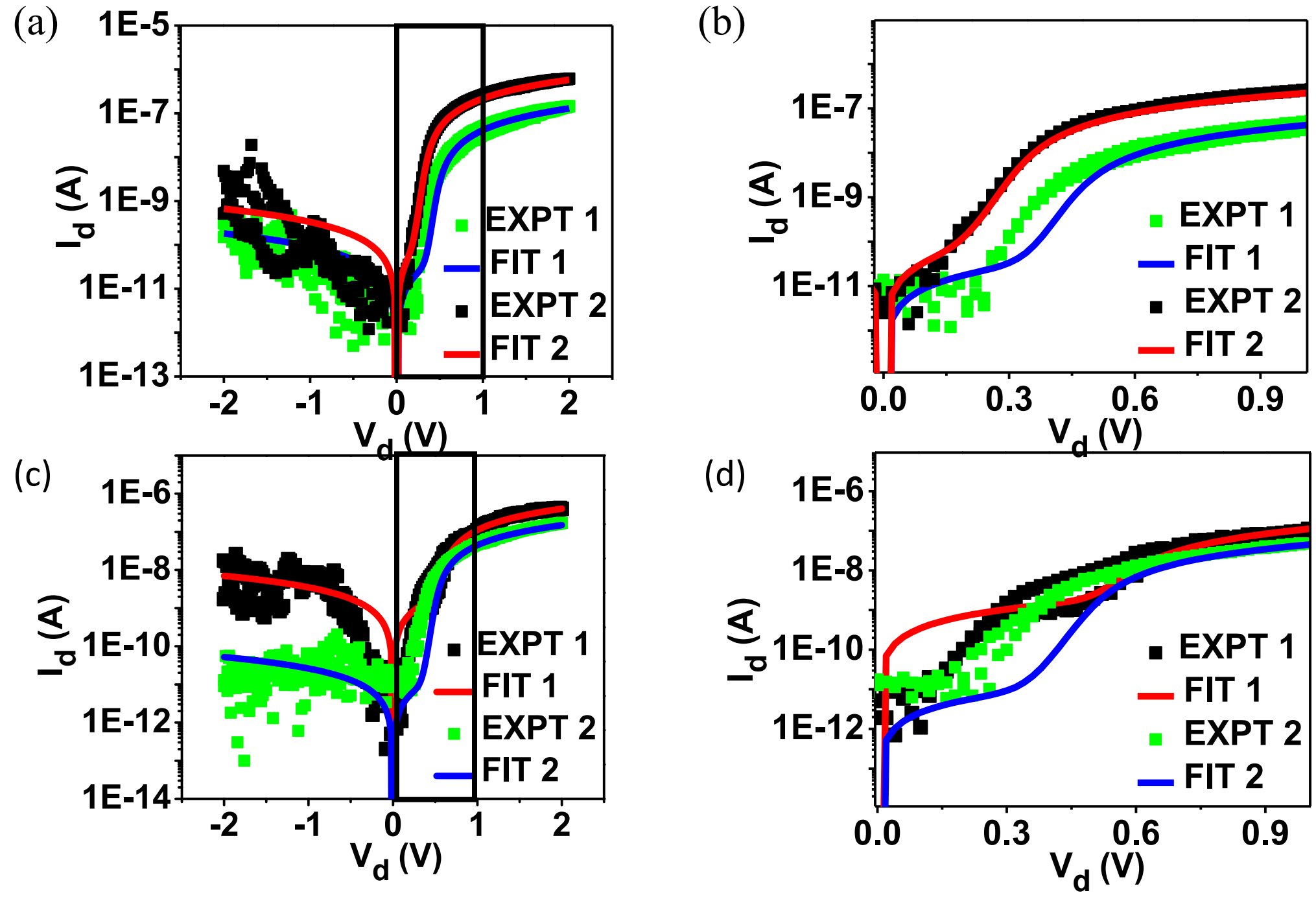

(a) I-V graphs of 2 single s-SWNT diodes that fit the model in figure 2a of main text well. Green and black curves represent the experimental I-V graphs of diode 1 and 2 respectively. Blue and red curves represent their respective modeled fit based on the equivalent circuit model shown in figure 2a of main text.

(b) This is a close-up of the boxed up region in part (a) to show the details of the good fit to the model between drive voltage of 0 to 1V.

(c) I-V graph of 2 single s-SWNT diodes that have less than ideal fit to the model in figure 2a of main text. Black and green curves represent the experimental I-V graphs of another 2 diodes, 1 and 2 respectively. Red and blue curves represent their respective modeled fit based on the equivalent circuit model shown in figure 2a of main text.

(d) This is a close-up of the boxed up region in part (c) to show the details of the less than ideal fit to the model between drive voltage of 0 to 1V.

Supplementary figure S3

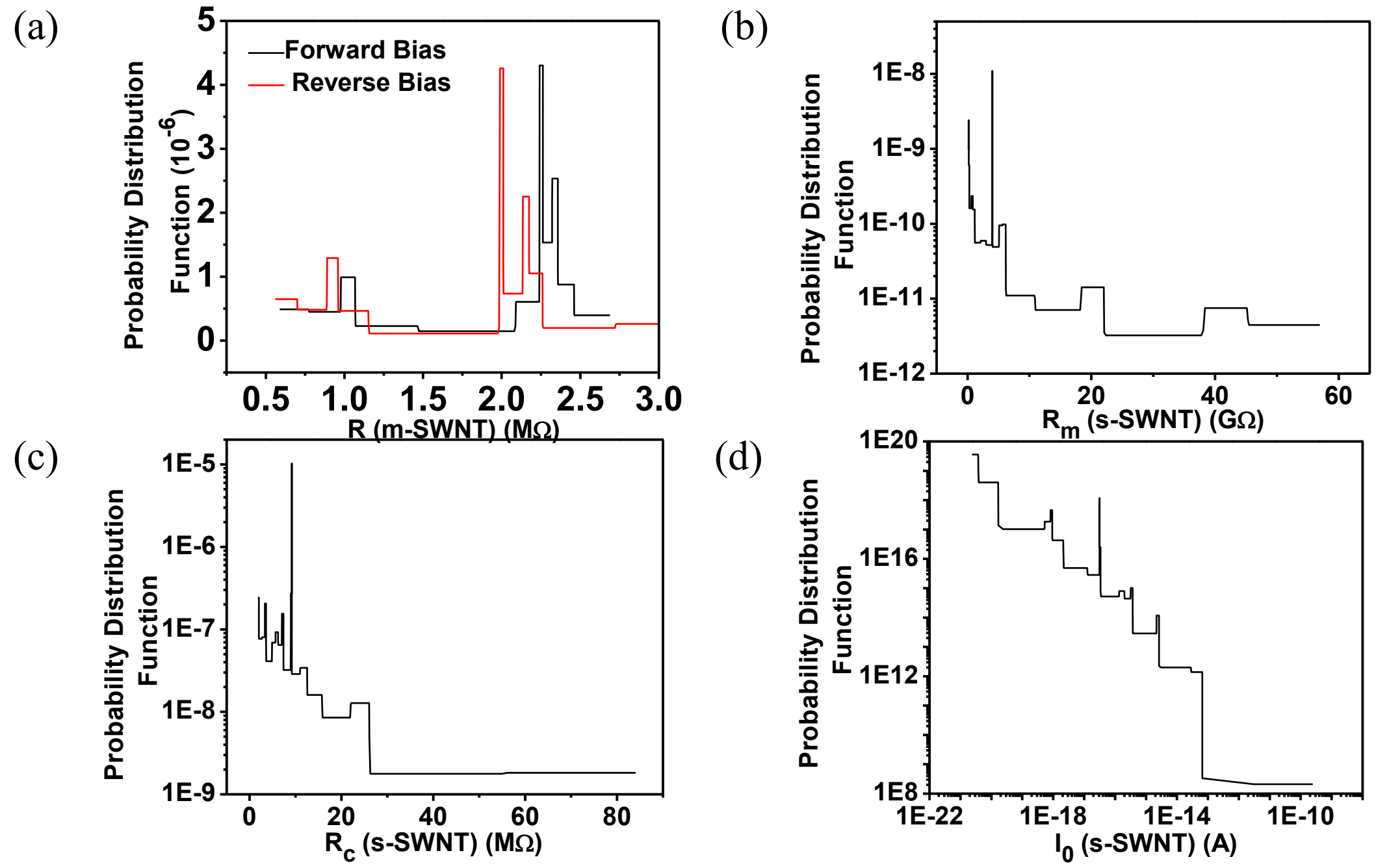

(a) Probability distribution function (PDF) for the resistance of the single m-SWNT resistors at forward (black curve) and reverse (red curve) biases.
(b) Probability distribution function (PDF) for $R_m$ of the single s-SWNT diodes.
(c) Probability distribution function (PDF) for $R_c$ of the single s-SWNT diodes.
(d) Probability distribution function (PDF) for $I_0$ of the single s-SWNT diodes.

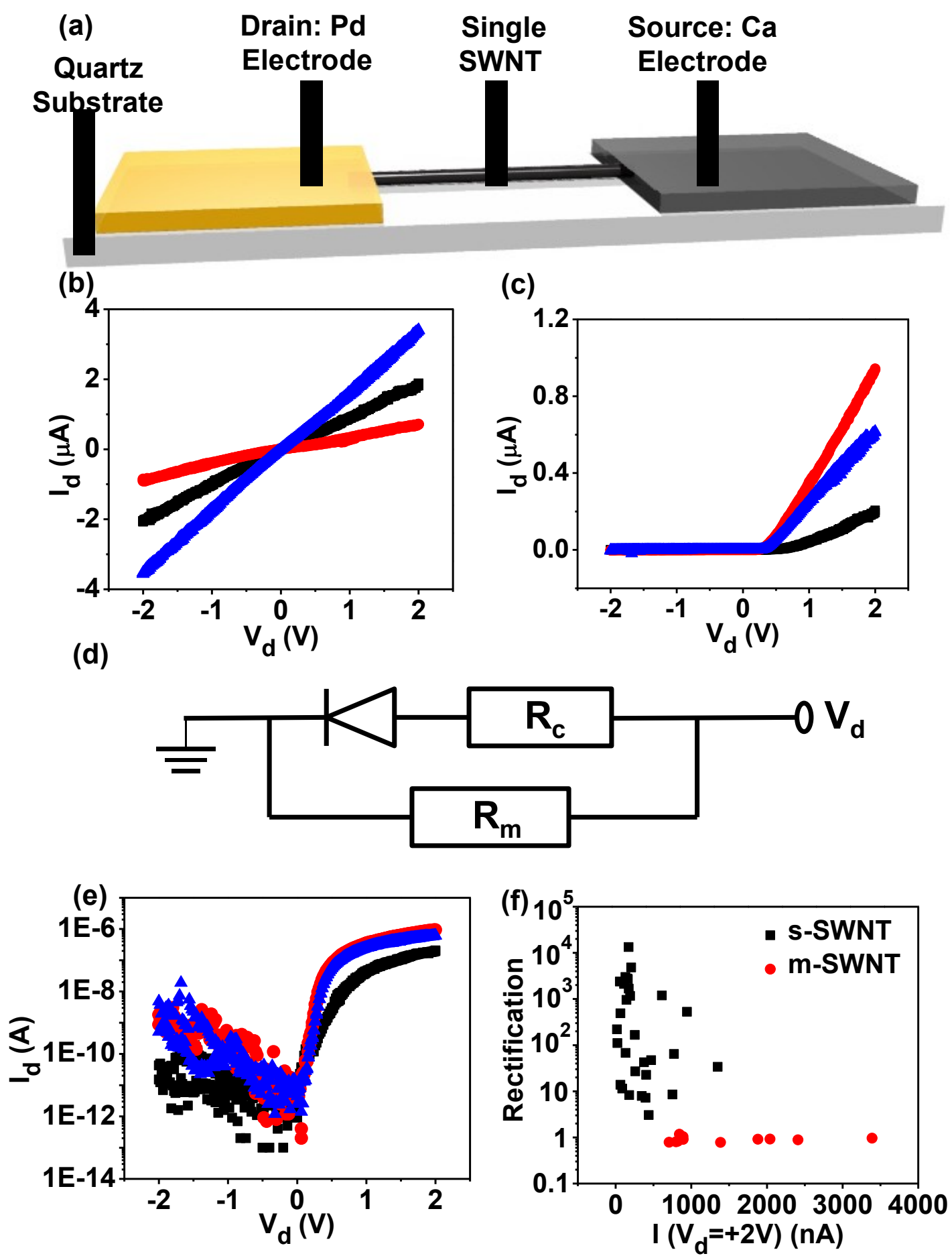

Figure 1:

FIG. 1. (a) Schematic illustration of a single single-walled carbon nanotube (SWNT) device. A single SWNT is contacted on one end by a palladium electrode and by a calcium electrode on the other end. (b) Three representative I-V curves (black, red and blue curves) of three single metallic-SWNT (m-SWNT) resistors. (c) Three representative I-V curves (black, red and blue curves) of three single semiconducting-SWNT (s-SWNT) diodes. (d) Equivalent circuit model of a non-ideal diode and a leakage via a parallel channel. *Rc* represents the Pd contact and channel resistance in series with the diode and *Rm* represents the shunt resistance that contributes to the leakage current. **(e)** Same I-V curves (black, red and blue curves) as in part (c) shown in log 10 scale. (f) Rectification (i.e. current at $V_d$=+2V divided by absolute value of current at $V_d$=-2V) as a function of the current at $V_d$=+2V for single s-SWNT diodes (black squares) and single m-SWNT resistors (red circles).

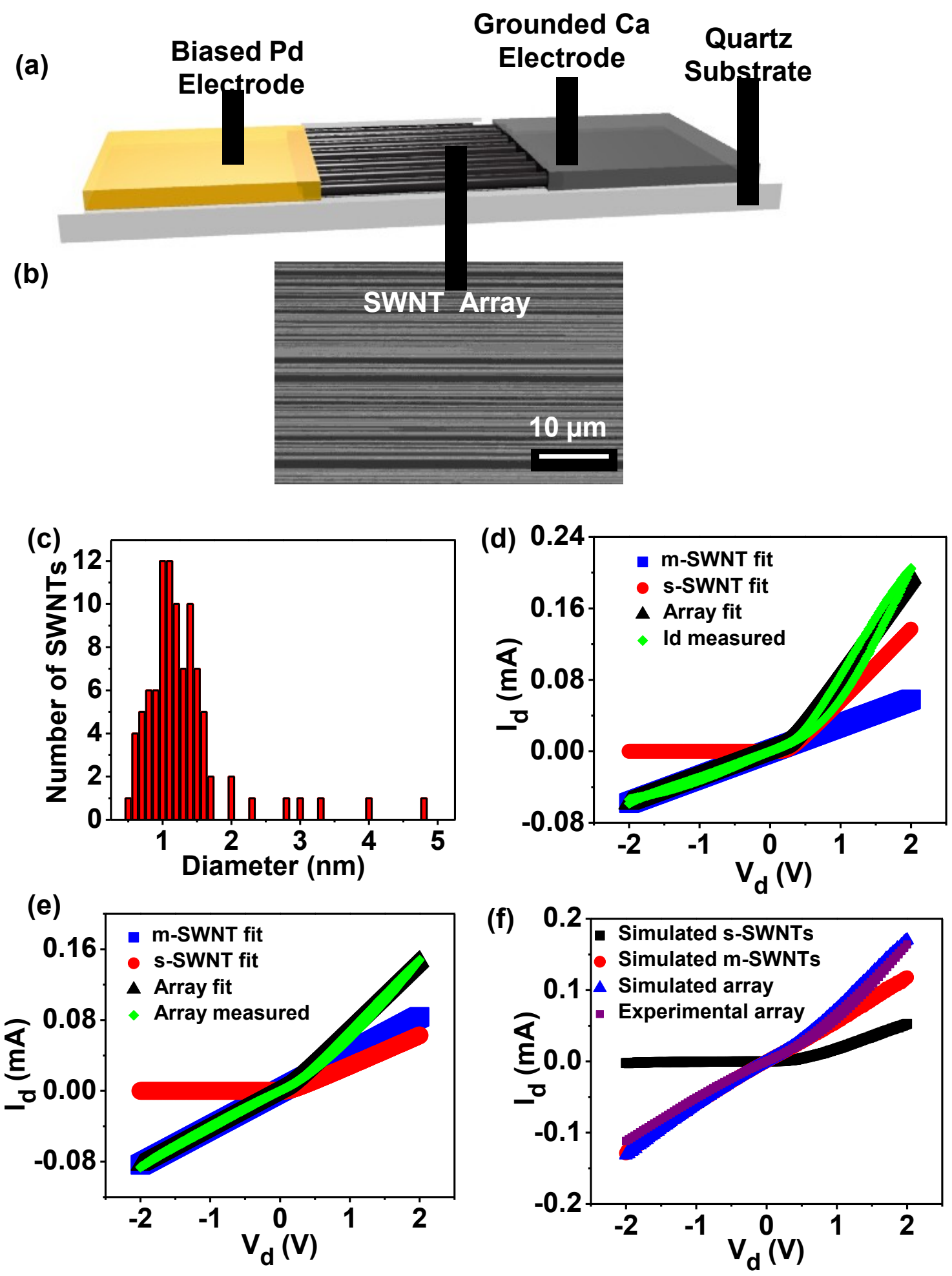

Figure 2:

FIG. 2. (a) Schematic illustration of a Schottky diode based on perfectly aligned arrays of SWNTs with an analogous design that in figure 1(a). (b) A SEM image of a representative array of SWNTs. (c) Diameter distribution of the SWNTs in the perfectly aligned arrays of SWNTs. (d) I-V curve of an array diode: model (black curve), including the current of the s-SWNTs (red curve) in the array is approximately twice that of the m-SWNTs (blue curve), and the corresponding measured data (green curve). (e) I-V curve of an array diode: model (black curve), including the current of the s-SWNTs (red curve) in the array is about the same as that of the m-SWNTs (blue curve), and the corresponding measured data (green curve). (f) Average current contributed by s-SWNTs (black curve) and m-SWNTs (red curve) in an array diode interpolated from the current of single SWNT devices. Blue curve represents the total average current of the array diode. The purple curve is the I-V curve of a measured array diode.

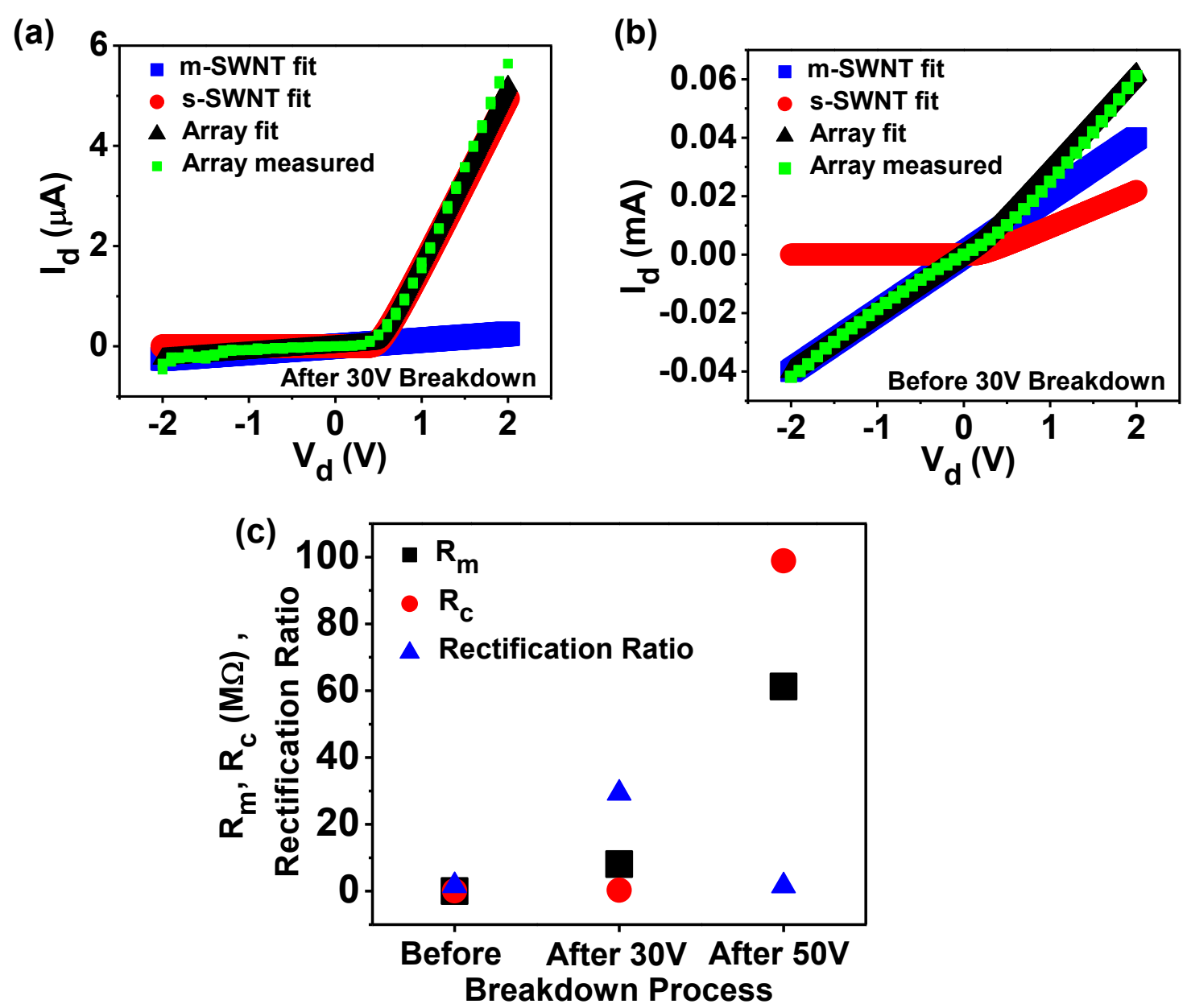

Figure 3:

FIG. 3. (a) I-V curve of an array diode after electrical breakdown by driving the device to $V_d$=30V: model (black curve), the current of the m-SWNTs (blue curve) is significantly lower than the s-SWNTs (red curve), and the corresponding measured data (green curve). (b) I-V curve of the same array diode as in part (a) before electrical breakdown: model (black curve) with its s-SWNTs (red curve) and m-SWNTs (blue curve) components and the corresponding measured data (green curve). (c) Plots comparing the values of $R_m$ (black square symbols), $R_c$ (red circle symbols) and rectification ratio (blue square symbols) before breakdown, after driving the device to $V_d$=30V and after driving the device to $V_d$=50V.